# Stock Price Prediction Using CNN and LSTM-Based Deep Learning Models

Sidra Mehtab
Department of Data Science
Praxis Business School
Kolkata, INDIA
email: smehtab@acm.org

Jaydip Sen
Department of Data Science
Praxis Business School
Kolkata, INDIA
email: jaydip.sen@acm.org

*Abstract*—Designing robust and accurate predictive models for stock price prediction has been an active area of research over a long time. While on one side, the supporters of the *efficient market hypothesis* claim that it is impossible to forecast stock prices accurately, many researchers believe otherwise. There exist propositions in the literature that have demonstrated that if properly designed and optimized, predictive models can very accurately and reliably predict future values of stock prices. This paper presents a suite of deep learning-based models for stock price prediction. We use the historical records of the NIFTY 50 index listed in the National Stock Exchange (NSE) of India, during the period from December 29, 2008 to July 31, 2020 for training and testing the models. Our proposition includes two regression models built on *convolutional neural networks* (CNNs), and three *long-and-short-term memory* (LSTM) network-based predictive models. For the purpose of forecasting the *open* values of the NIFTY 50 index records, we adopted a multi-step prediction technique with *walk-forward validation*. In this approach, the *open* values of the NIFTY 50 index are predicted on a time horizon of one week, and once a week is over, the actual index values are included in the training set before the model is trained again, and the forecasts for the next week are made. We present detailed results on the forecasting accuracies for all our proposed models. The results show that while all the models are very accurate in forecasting the NIFTY 50 *open* values, the univariate encoder-decoder convolutional LSTM with previous two weeks' data as the input is the most accurate model. On the other hand, a univariate CNN model with previous one week's data as the input is found to be the fastest model in terms of its execution speed.

I. INTRODUCTION

Analysis of financial time series and prediction of future stock price values and future stock price movements have been an active area of research over a long period of time. While there are researchers who believe in the well-known efficient market hypothesis, and claim that it is impossible to forecast stock prices accurately, propositions exist in the literature that demonstrate that it is possible to predict the values of stock prices with a very high level of accuracy using carefully designed predictive models. It has also been found that the accuracy of a predictive model depends on the set of variables used in building the model, the algorithms deployed, and how the model has been optimized. There are propositions in the literature that focus on the decomposition of time series for stock price prediction [1-2]. Applications of machine learning and deep-learning approaches have also been quite popular in stock price movement analysis and forecasting [3-4].

Mehtab and Sen propose a model for stock price forecasting that utilizes the sentiment of the investors from the social media in augmenting the output of a deep learning framework to arrive at a very high level of accuracy in prediction. The proposed framework also deploys a non-linear multivariate system built on a *self-organizing fuzzy neural network* (SOFNN) [5]. In two recently published work, Mehtab and Sen presented a suite of *convolutional neural network* (CNN)-based regression models that exhibited a very high level of accuracy and robustness in forecasting on a multivariate financial time series data [6-7].

Several propositions exist in the literature on technical analysis of stock price movement patterns. Among the various indicators of price movements, *moving average divergence* (MACD), *momentum stochastics*, *meta sine wave*, etc. are quite well known. These indicators provide the investors with a rich set of visualization platforms and useful metric that help investors in making effective decisions on investment in the stock market.

In this work, we propose a suite of deep learning-based regression models for the purpose of forecasting NIFTY 50 index values. For building the models, the historical values of the NIFTY 50 index for the period December 29, 2008 (which was a Monday) to December 28, 2018 (which was a Friday) have been used as the training records. The models have been tested on NIFTY 50 index values during the period of December 31, 2018 (which was a Monday) to July 31,2020 (which was a Friday). The five models that we propose in this work include two *convolutional neural network* (CNN)-based models, and three *long- and short-term memory* (LSTM) network-based models. The models have different architectures and different structures in their input data. While all the models have univariate input data, four of them use the previous two weeks' data as their input for forecasting the *open* values of the NIFTY 50 index time series. However, one CNN model uses the previous one week's data as the input for the purpose of forecasting the *open* value of the NIFTY 50 index of the next week.

The organization of the paper is as follows. In Section II, we present a clear definition of the problem we solve in this paper. Section III presents a very brief outline on some related work in the field of stock price forecasting. In Section IV, we describe the methodology followed by us in this work. This section also presents the architectural details of all our proposed models. The results on the performance of the models are presented in Section V. Finally, in Section VI, we conclude the paper while highlighting some future research directions.

## II. PROBLEM STATEMENT

Our objective is to build a robust and accurate predictive framework that contains a suite of deep learning-based regression models. We have used the historical records of NIFTY 50 index values over a period of five and a half years for building and testing our proposed models. We have chosen a very realistic value of the prediction horizon as one week for our proposed models. We hypothesize that the deep learning models will be able to extract a rich feature set from the past NIFTY 50 index values and will be able to forecast the future index values with a very high level of accuracy. In our past work, we proposed a suite of four CNN-based regression models to validate our hypothesis [6]. In the current work, we augment our proposition with five different deep learning-based regression models. While two of the proposed models are built on CNN, the remaining three models are based on three variants of LSTM network architecture.

## III. RELATED WORK

Design and development of models for forecasting of stock prices and movement of stock prices have been a very active area of research. While extensive work has been done on these areas, most of the existing propositions in the literature can be categorized into three broad types. The frameworks belonging to the first category are essentially built one multivariate ordinary least square regression [8-10]. However, these models fail to perform well on real-world data as the stringent requirements that these models impose on the data are usually not satisfied. The propositions in the second category are time series and econometric models like *autoregressive moving average* (ARIMA), *Granger causality*, *quantile regression* etc. [11-13]. These models yield high level of accuracy in forecasting if the financial time series data is largely dominated by trend and a seasonal component. However, their accuracy level falls drastically in presence of any strong random component in the time series. The predictive models of the third category are based on machine learning, deep learning, and natural language processing algorithms [14-17]. These models learn from the patterns in the past data and the textual information in the web and social media, and exploit that information in forecasting future stock prices. Performance of the models has been found to be superior on financial time series data in comparison to the models of the first two categories.

Most of the existing propositions in the literature on stock price prediction suffer from a common shortcoming. If the stock price time series exhibits significant randomness, the forecast accuracies of the models drastically decrease. The proposed models in our current work have yielded very high level of accuracies by utilizing the power of *convolutional neural networks* (CNNs) and *long-and-short-term memory* (LSTM) networks in their ability in learning deep features from the past values of a financial time series. The learned features are used for making forecasts for the future values of the stock index. Moreover, time needed for execution of the models were found to be quite moderate on our target hardware architecture. The fastest model in our proposition needed only 11.17s on an average, for model construction using a training dataset consisting of 1045 records and testing it on a test dataset consisting of 415 records.

## IV. METHODOLOGY

As we mentioned in Section II, the main objective of this work is to build a suite of predictive framework for accurately forecasting the daily values of NIFTY 50 index. For training and testing our proposed predictive models, we use the historical NIFTY index values for the period during December 29, 2014 to July 31, 2020. The NIFTY index records were downloaded in the form of a *comma separated variables* (CSV) file from the Yahoo Finance website [18]. The following attributes constituted the daily records of NIFTY 50 index values: (i) *date*, (ii) *open*, (iii) *high*, (iv) *low*, (v) *close*, and (vi) *volume*.

The predictive models proposed in this work are all deep learning-based regression models. We use the variable *open* as the *response* variable, and all the other variables are used as the predictors. NIFTY 50 daily data for the period December 29, 2014 to December 28, 2018 has been used as the training data for building the models, while we tested the models using the data for the period December 31, 2018 to July 31, 2020. Hence, the training dataset comprised of 1045 records spanning over 209 weeks, while the test dataset consisted of 415 records over 83 weeks. We followed the approach of *multi-step forecasting with walk-forward validation* for the purpose of validation and testing of our proposed models [19]. Using this method, we build the models based on the training dataset and forecast the *open* values of the NIFTY 50 index on weekly basis for the records in the test dataset. As a week gets over, the actual *open* values of the records for that week are included in the training dataset and forecasting for the *open* values for the next week is done. NSE of India remains operational for five days a week – Monday to Friday. Hence, each round of forecasting involves forecasting of the *open* values corresponding to those five days in the upcoming week.

To make our forecasting framework more robust and accurate, we build some deep learning-based regression models too. In one of our previous work, we demonstrated the effectiveness and accuracy of *convolutional neural networks* (CNNs) in forecasting time series index values [6]. In this work, in addition to exploiting the power of CNN, we have utilized another type of deep learning model - long-*and-short-term memory* (LSTM) networks - in forecasting on a complex multivariate time series like the NIFTY 50 series.

A CNN consists of two major processing layers – the *convolutional* layers and the *pooling* layers [7]. The convolutional layers are used for reading the inputs either in the form of a two-dimensional image or as a sequence of one-dimensional data. The results of the reading are projected into a filter map that represents the interpretation of the input. The pooling layers operate on the extracted feature maps and derives the most essential features by averaging (average pool) or max computing (max pooling) operations. For extracting deep features from the input sequence, the convolution and the pooling layers may be repeated multiple times. The output from the last pooling layer is sent to a one or more *dense* layer(s) for extensive learning from the input data.

LSTM is a deep neural network architecture that essentially belongs to the family of *recurrent neural networks* (RNNs). RNNs have a characteristic that distinguishes these networks from other deep neural networks – they have feedback loops [19]. However, RNNs suffer from a problem known as the *vanishing and exploding gradient problem*, in which the network either stops learning or continues to learn at a very high *learning rate* so that it never converges to the point of the minimum error. The architectures of LSTM networks are designed in such a way that the problem of vanishing or exploding gradient never occur in these networks, and hence, such networks are found to be very suitable in modelling complex sequential data such as texts and time series. These networks consist of *cells* that store historical state information of the network, and *gates* that regulate and control the flow of information through these cells. Three types of gates are used in an LSTM network – *forget gates*, *input gates*, and *output gates*. The forget gates are instrumental in throwing away irrelevant past information, and in remembering only the information that is relevant for the current slot. The input gates control the new information that acts as the input to the current state of the network. The memory cells in the network intelligently aggregate the old state information from the forget gates and the current input to the network received through the input gate. Finally, the output gates produce the output from the network at the time slot. The output can be considered as the forecasted value computed by the model for the current slot [19].

In this paper, we have presented five different predictive models. The models are different in their architectures and their input data shapes are also dissimilar. The five models are: (i) CNN model with univariate input data of past one week, (ii) CNN model with univariate input data of past two weeks, (iii) Encoder-decoder LSTM with univariate data of past two weeks, (iv) Encoder-decoder CNN LSTM model with univariate input data of past two weeks, and (v) Encoder-decoder Convolutional LSTM with univariate input data of past two weeks.

In the following, we briefly discuss the salient architectural details of the two CNN and three LSTM models that we have proposed in this work.

The first model is a univariate CNN model that uses the previous one week's data as the input and carries out multi-step forecasting with walk-forward validation. The model is trained using the records in the training dataset, and then it is used to forecast the *open* values for the five days in the next week. The shape (5, 1) of the input data to the network refers to only one attribute, i.e., the *open* values, of the five days in the previous week. The CNN model consists of only one convolution layer that extracts 16 feature maps from the input data with a kernel of size 3. The *convolution layer* enables the network to read the input data of five days in three time-steps with each reading resulting in the extraction of 16 features. The *subsampling layer* following the convolutional layer performs a *max-pooling* operation of size 2, thereby reducing the size of the feature maps. The output of the subsampling layer is then converted into a one-dimensional vector and then interpreted by a *fully-connected layer* before the output layer (which is also fully-connected) predicts the *open* values for the next five days. The *rectified linear unit* (ReLU) function has been used in the convolution and fully-connected layer. The performance of the layers is optimized by using the ADAM version of the *stochastic gradient descent algorithm*. We trained the model using 20 epochs with each batch consisting of 4 input values. For measuring the level of accuracy in forecasting of the models, we have used *root mean square error* (RMSE) as the metric. Fig. 1 depicts the architecture of the CNN model for *Case I*. Throughout this paper, we will refer to this model as CNN#1.

Our second model is also a CNN-based regression model that uses previous two week's *open* values as the univariate input. This model has exactly the same architecture and same parameters as those of the CNN#1 model. However, unlike CNN#1, the model, in this case, is fed with the previous two weeks' data (i.e., 10 records as the data input). We refer to this model as CNN#2, whose architecture is presented in Fig. 2.

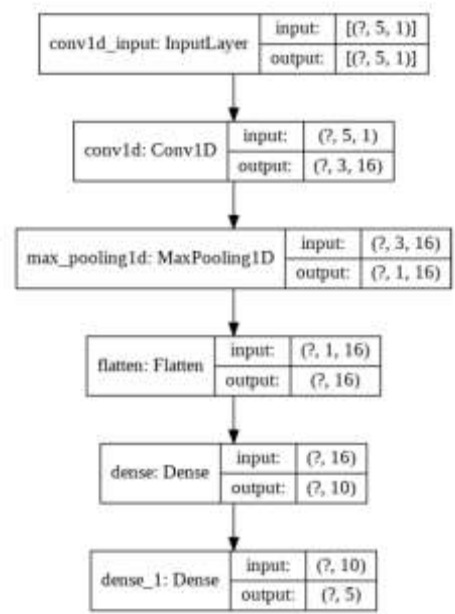

Fig. 1. The architecture of univariate CNN model (CNN#1) with prior one week's data as the input

The third model in the suite is a univariate LSTM model that consists of an *encoder* and a *decoder* sub-model and uses the previous two weeks data as the input. The *encoder sub-model* is used for reading and encoding the input sequence, while the *decoder sub-model* is delegated with the responsibility of interpreting the encoded input sequence, and then making a one-step prediction at a time. The decoder sub-model of the model uses an LSTM network that allows it to memorize the values that was predicted in the previous round (i.e., for the previous day) in the output sequence, and store the internal state of the network while producing the predicted values in the output sequence. The first LSTM layer that acts as a decoder sub-model, and it consists of 200 nodes. This layer reads the input sequence with shape (10, 1), indicating only one attribute (i.e., the open value) for the previous two weeks is fed as the input to the layer. The output of the LSTM decoding layer is a 200-element vector (one element per node) that extracts deep features from the 10 input values. For each time step in the output sequence, the input data sequence is analyzed once. With five time-steps and 200 features extracted by the LSTM decoding layer, the repeat vector layer takes the shape (5, 200). The output sequence is now decoded by another decoder LSTM model with 200 units [19]. The output sequence is then interpreted in each time-step by a full-connected layer before it is sent to the final output layer. The final prediction is done in a step-by-step manner, and not for all the five days in a week at a time. Essentially it implies that the same fully connected layer and output layer are being utilized for each step of forecasting. This is implemented using a *TimeDistributed wrapper* layer that creates a wrapped layer to be used for each time-step of the forecasted sequence [19]. As the model predicts one value at each time-step (i.e., prediction for a single day), the weekly prediction will have a shape (5,1).

extracts 64 feature maps from the input sequence with a kernel size of three time-stamps. The second convolutional layer performs the same operation on the feature maps produced by the first and amplifies the features. A *max-pooling* layer follows the second convolutional layer. The max-pooling layer reduces the feature maps by half by retaining the maximum values of the features. The output feature maps values of the max-pooling layer are then flattened into a long vector consisting of 192 values as depicted in Fig. 4. This flattened vector is fed as an input to the decoder LSTM sub-model. The decoder LSTM sub-model remains exactly identical to that in the LSTM#1 model discussed earlier. As in the case of LSTM#1, this model is also trained using a batch size of 16, over 20 epochs. The architecture of the model is depicted in Fig. 4. We refer to the is model as the LSTM#2 model.

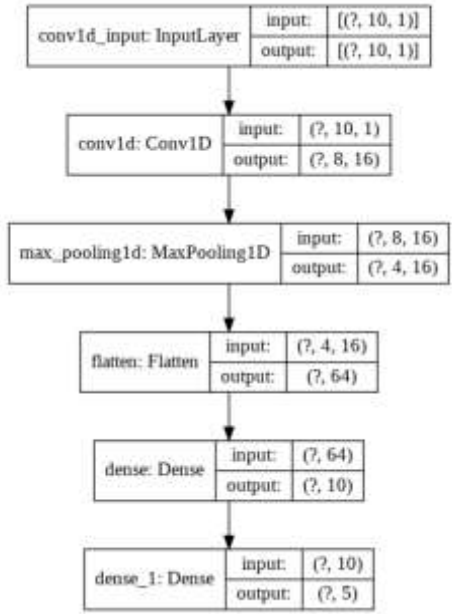

Fig. 2. The architecture of univariate CNN model (CNN#2) with prior two weeks' data as the input

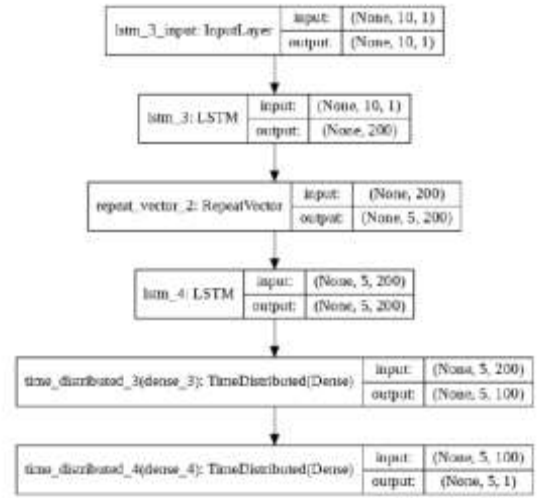

Fig. 3. The architecture of univariate encoder-decoder LSTM model (LSTM#1) with prior two weeks' data as the input

We have used the ReLU activation function in the two decoder LSTM layers and the *TimeDistributed Wrapper* layer, and the loss function and the optimizer used in the output layer were MSE and ADAM respectively. The model is trained over 20 epochs using a batch size of 16. These are the optimum values of the hyperparameters obtained using the grid search technique. We refer to this model as the LSTM#1 whose architecture is presented in Fig. 3.

The fourth model and the second one in the LSTM family that we propose is a variant of a *univariate encoder-decoder* LSTM model that uses a CNN in the encoder layer. While CNNs do not directly support the interpretation of any sequential data, we have utilized the capability of a one-dimensional CNN in reading sequential data and extracting important features from it. The features extracted by the CNN are then fed into an LSTM for decoding and forecasting of a sequential series. Since the model uses a CNN as the encoder and LSTM as the decoder, we call this model a CNN-LSTM model. In this work, we have built a univariate CNN-LSTM model in which the encoder CNN comprises of two convolutional layers. The input to the model is the sequential series consisting of the *open* values of the previous two weeks. The first convolutional layer

The final model and the third one from the LSTM suite in our proposition is an extension of the CNN-LSTM model that performs the convolutions of the CNN (i.e., the way a CNN reads and extracts features from a sequential input data) as an integrated operation with the LSTM for each time-step. We refer to this variant as a *Convolutional LSTM* model [20]. A pure LSTM-based regression model reads the input sequential data directly, and computes the internal state and state transitions to produce its predicted sequence. On the other hand, a CNN-LSTM model extracts the features from the input sequence using a CNN encoder sub-model, and then interprets the output from the CNN models, using the LSTM decoder sub-model. In contrast to these, a Convolutional LSTM model uses convolution operation to directly read a sequential input into a decoder LSTM sub-model. We have reconfigured the ConvLSTM2d class in Keras library that supports two-dimensional ConvLSTM models so that it can handle a one-dimensional univariate time series. With two weeks' univariate time series values as the input, the sequence can be visualized as one row with 10 columns. Although a Convolutional LSTM can read this sequence in a single time-stamp, a single reading and a subsequent convolution operation are not sufficient for deep feature extraction from the input sequence. Hence, we divide the sequential data for 10 days into two subsequences, each with a length of 5 days. The Convolutional LSTM model reads the two subsequences in two time-steps and performs a convolution operation after each time-step. This enables the model to extract more features from the input sequence. The input data in the training set was reshaped into the following structure: [*sample no.*, *time-step*, *row*, *column*, *channels*]. In our model design, the values of the time-step, row, column, and channels were 2, 1, 5, and 1 respectively as considered only a univariate sequence consisting of the *open* values only. The schematic architecture of the model is represented in Fig. 5. We refer to this model as the LSTM#3 model.

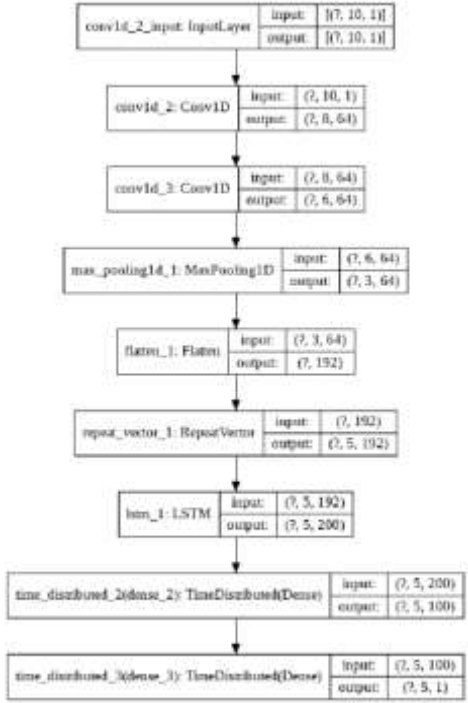

Fig. 4. The architecture of the univariate encoder-decoder CNN-LSTM model with prior two weeks' data as input (LSTM#2)

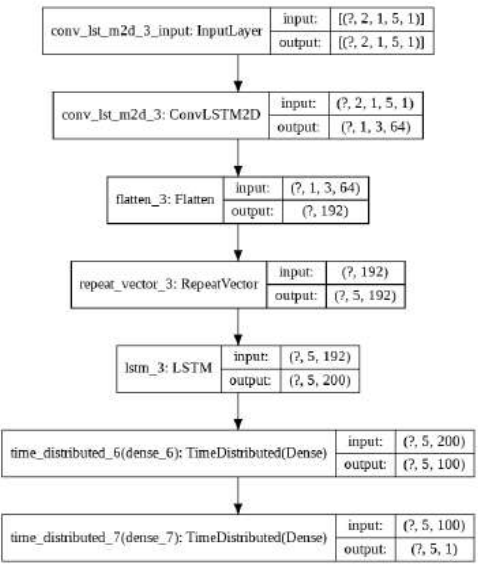

Fig. 5. The architecture of the univariate encoder-decoder Convolutional LSTM model with prior two weeks' data as the input (LSTM#3)

## V. PERFORMANCE RESULTS

In this Section, we present the performance results of the deep learning models. The details of the design of all the five models were presented in Section IV. We tested each model for 10 rounds and noted their performances in terms of the overall RMSE, the RMSE values for the individual days of a week (i.e., Monday – Friday), the execution time of the model, and the ratio of the RMSE of the model to the mean of the *open* values of the records in the test dataset. It may be noted that the number of records in the training and the test dataset was 1045 and 415 respectively. The mean *open* value in the test dataset was 11070.59.

TABLE I. RESULTS OF UNIVARIATE CNN MODEL WITH PREVIOUS ONE WEEK'S DATA AS INPUT (CNN#1)

| No. | RMSE | Mon | Tue | Wed | Thu | Fri | Time |
|---|---|---|---|---|---|---|---|
| 1 | 379.7 | 272 | 331 | 370 | 431 | 464 | 11.96 |
| 2 | 403.7 | 308 | 353 | 405 | 445 | 484 | 13.06 |
| 3 | 382.0 | 276 | 332 | 381 | 430 | 461 | 12.68 |
| 4 | 393.1 | 289 | 346 | 394 | 440 | 469 | 8.60 |
| 5 | 379.4 | 268 | 335 | 375 | 422 | 465 | 13.39 |
| 6 | 382.3 | 272 | 357 | 371 | 427 | 458 | 12.58 |
| 7 | 385.2 | 286 | 334 | 372 | 444 | 462 | 12.88 |
| 8 | 370.1 | 264 | 329 | 360 | 415 | 453 | 12.61 |
| 9 | 399.2 | 317 | 334 | 408 | 447 | 469 | 14.58 |
| 10 | 390.1 | 284 | 337 | 387 | 449 | 464 | 12.49 |
| **Mean** | **386.47** | 283 | 339 | 382 | 435 | 465 | **12.48** |
| **Min** | 370.1 | 264 | 329 | 360 | 415 | 453 | 8.60 |
| **Max** | 403.7 | 317 | 357 | 408 | 449 | 484 | 14.58 |
| **SD** | 10.11 | 17.2 | 9.65 | 15.8 | 11.5 | 8.2 | 1.53 |
| **RMSE/Mean** | **0.0349** | 0.03 | 0.03 | 0.03 | 0.04 | 0.04 | |

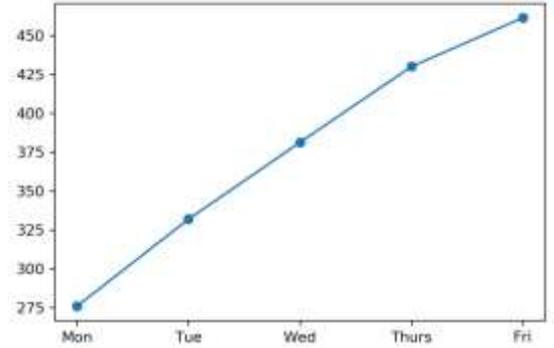

Fig. 6. Variation of RMSE with different days in a week for the CNN#1 model (Round #3 of Table I)

Table I presents the performance results of the CNN#1 model. The model has been executed on a system consisting of an Intel i5-8250U processor with clock speed 1.60 GHz – 1.80 GHz, 8 GB RAM, and running 64-bit Windows 10 operating system. The execution time has been measured in seconds. The model was evaluated over 10 rounds and it is observed that the CNN#1 model needed an average of 11.92s for the execution of one epoch. The mean value of the ratio of the RMSE to the mean of the *open* values in the test dataset records produced by the model was 0.0349. It is also interesting to note that the mean RMSE values consistently increased from Monday through Friday. Fig. 6 presents the performance results of the CNN#1 model for round #3 presented in Table I.

Table II presents the performance of the CNN#2 regression model. The mean execution time for the 10 rounds of execution of the model on the same computing environment was found to be 13.87s. This was just a little higher than the time needed for the execution of the CNN#1 model. The model exhibited an average value of 0.0341 as the ratio of the RMSE to the mean *open* values in the test dataset records. Hence, in terms of both the metrics – RMSE to mean *open* value, and the mean execution time – CNN#2 model is found to be inferior to the CNN#1 model. Fig. 7 presents the performance results of the CNN#2 model for round #7 presented in Table II.

TABLE II. RESULTS OF UNIVARIATE CNN MODEL WITH PREVIOUS TWO WEEKS' DATA AS INPUT (CNN#2)

| No. | RMSE | Mon | Tue | Wed | Thu | Fri | Time |
|---|---|---|---|---|---|---|---|
| 1 | 404.7 | 305 | 359 | 397 | 449 | 488 | 12.64 |
| 2 | 417.2 | 358 | 346 | 385 | 478 | 494 | 16.84 |
| 3 | 427.7 | 332 | 404 | 412 | 463 | 507 | 13.38 |
| 4 | 422.3 | 314 | 419 | 405 | 455 | 497 | 12.89 |
| 5 | 416.0 | 329 | 370 | 407 | 460 | 493 | 14.12 |
| 6 | 424.5 | 334 | 381 | 418 | 466 | 502 | 14.53 |
| 7 | 402.9 | 307 | 355 | 391 | 430 | 504 | 13.46 |
| 8 | 421.3 | 335 | 378 | 414 | 463 | 497 | 13.58 |
| 9 | 435.5 | 379 | 401 | 426 | 463 | 497 | 13.32 |
| 10 | 460.0 | 382 | 421 | 448 | 498 | 535 | 13.90 |
| **Mean** | **423.23** | 338 | 383 | 410 | 463 | 501 | **13.87** |
| **Min** | 402.9 | 305 | 346 | 385 | 430 | 488 | 12.64 |
| **Max** | 460.0 | 382 | 421 | 448 | 498 | 535 | 16.84 |
| **SD** | 16.23 | 27.3 | 26.6 | 18.4 | 17.9 | 13.1 | 1.18 |
| **RMSE/Mean** | **0.0382** | 0.03 | 0.03 | 0.04 | 0.04 | 0.05 | |

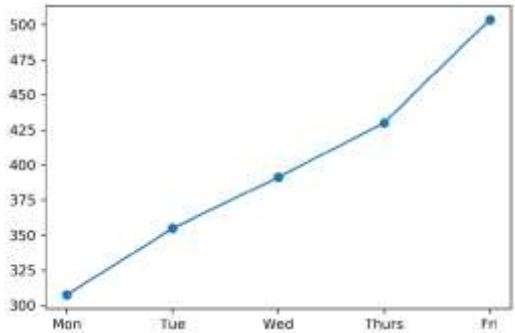

Fig. 7. Variation of RMSE with different days in a week for the CNN#2 model (Round #7 of Table II)

TABLE III. RESULTS OF UNIVARIATE ENCODER-DECODER LSTM MODEL WITH PREVIOUS TWO WEEKS' DATA AS INPUT (LSTM#1)

| No. | RMSE | Mon | Tue | Wed | Thu | Fri | Time |
|---|---|---|---|---|---|---|---|
| 1 | 391.7 | 318 | 383 | 395 | 433 | 418 | 12.79 |
| 2 | 418.1 | 367 | 398 | 415 | 459 | 446 | 12.56 |
| 3 | 409.1 | 334 | 381 | 403 | 462 | 452 | 14.95 |
| 4 | 423.0 | 365 | 400 | 413 | 467 | 461 | 14.74 |
| 5 | 403.4 | 326 | 414 | 389 | 424 | 453 | 14.79 |
| 6 | 397.9 | 349 | 379 | 393 | 440 | 422 | 14.68 |
| 7 | 389.8 | 344 | 384 | 372 | 425 | 418 | 15.11 |
| 8 | 395.6 | 327 | 362 | 391 | 445 | 440 | 15.44 |
| 9 | 449.0 | 343 | 387 | 468 | 527 | 493 | 14.95 |
| 10 | 412.1 | 348 | 382 | 411 | 456 | 453 | 15.26 |
| **Mean** | **408.97** | 342 | 387 | 405 | 454 | 446 | **14.53** |
| **Min** | 390 | 318 | 362 | 372 | 424 | 418 | 12.56 |
| **Max** | 449 | 367 | 414 | 468 | 527 | 493 | 15.44 |
| **SD** | 17.92 | 16.2 | 14.0 | 26.0 | 29.3 | 22.9 | 1.00 |
| **RMSE/Mean** | **0.0369** | 0.03 | 0.04 | 0.04 | 0.04 | 0.04 | |

Table III presents the performance of the LSTM#1 model, which is a univariate encoder-decoder LSTM model. The model took 14.53s on average to execute over 10 rounds on our hardware system. The average value the ratio of the RMSE to the mean of the actual *open* values yielded by the model was 0.0369. Hence, in terms of both the metrics the LSTM#1 model was found to be inferior to the CNN#1 model. Fig. 8 depicts the behavior of the model RMSE over different days in the week as per the round #1 in Table III. It is evident from Table III, that model exhibited the highest RMSE on Thursdays.

Table IV presents the performance results of model #LSTM2 – a univariate encoder-decoder CNN-LSTM model. The mean time for execution of the model was 15.25s, and the average value the ratio of the RMSE to the mean of the actual *open* values yielded by the model was 0.0416. Fig. 9 shows the RMSE of the model with different days in the week as per round #2 in Table IV. It is evident from Table IV, that the mean RMSE of the model consistently increased from Monday to Friday in a week.

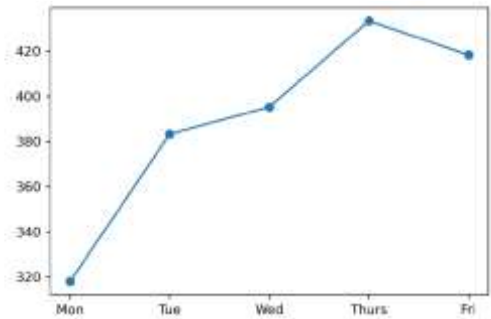

Fig. 8. RMSE of LSTM#1- univariate encoder-decoder LSTM time series with the previous one week's data as input (Round #1 of Table III)

TABLE IV. RESULTS OF UNIVARIATE ENCODER-DECODER CNN LSTM MODEL WITH PREVIOUS TWO WEEKS DATA AS INPUT (LSTM#2)

| No. | RMSE | Mon | Tue | Wed | Thu | Fri | Time |
|---|---|---|---|---|---|---|---|
| 1 | 535.8 | 389 | 486 | 523 | 602 | 642 | 15.03 |
| 2 | 477.1 | 353 | 420 | 463 | 533 | 583 | 16.29 |
| 3 | 451.0 | 344 | 409 | 452 | 496 | 531 | 15.14 |
| 4 | 458.1 | 358 | 413 | 452 | 502 | 543 | 14.96 |
| 5 | 532.5 | 389 | 483 | 634 | 549 | 574 | 15.22 |
| 6 | 436.9 | 338 | 390 | 432 | 480 | 521 | 15.28 |
| 7 | 438.5 | 343 | 396 | 428 | 480 | 521 | 15.51 |
| 8 | 402.6 | 314 | 359 | 397 | 443 | 479 | 14.97 |
| 9 | 416.8 | 327 | 375 | 410 | 459 | 493 | 14.17 |
| 10 | 451.8 | 342 | 402 | 439 | 502 | 546 | 15.92 |
| **Mean** | **460.11** | 350 | 413 | 463 | 505 | 543 | **15.25** |
| **Min** | 402.6 | 314 | 359 | 397 | 443 | 479 | 14.17 |
| **Max** | 535.8 | 389 | 486 | 634 | 602 | 642 | 16.29 |
| **SD** | 44.23 | 24.0 | 41.7 | 69.3 | 46.4 | 47.3 | 0.58 |
| **RMSE/Mean** | **0.0416** | 0.03 | 0.04 | 0.04 | 0.05 | 0.05 | |

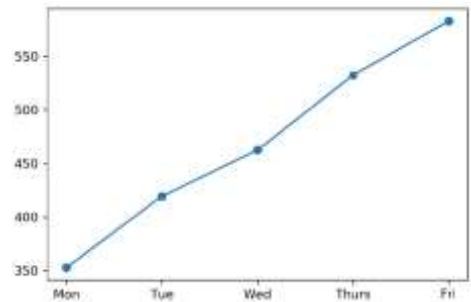

Fig. 9. Variation of RMSE with different days in a week for the LSTM#2-model (Round #3 of Table IV)

The performance of the model LSTM#3 is presented in Table V. This is a univariate encoder-decoder convolutional LSTM model with previous two weeks' data as its input. The mean execution time for the model for the 10 rounds was found to be 11.17s, while the ratio of the RMSE to the mean open value in the test dataset was 0.0350. Fig.10 depicts how the model RMSE varied with different days in a week as per the record # 5 in Table V.

TABLE V. RESULTS OF UNIVARIATE ENCODER-DECODER CONV. LSTM MODEL WITH PREVIOUS TWO WEEKS' DATA AS INPUT (LSTM#3)

| No. | RMSE | Mon | Tue | Wed | Thu | Fri | Time |
|---|---|---|---|---|---|---|---|
| 1 | 323.7 | 248 | 292 | 315 | 375 | 372 | 11.71 |
| 2 | 422.5 | 349 | 395 | 414 | 475 | 467 | 10.84 |
| 3 | 412.4 | 299 | 386 | 418 | 468 | 467 | 11.16 |
| 4 | 373.1 | 294 | 393 | 358 | 409 | 401 | 10.78 |
| 5 | 375.9 | 291 | 336 | 362 | 430 | 439 | 11.58 |
| 6 | 358.6 | 300 | 333 | 354 | 404 | 392 | 11.24 |
| 7 | 367.9 | 353 | 338 | 351 | 405 | 389 | 11.05 |
| 8 | 495.0 | 342 | 415 | 470 | 577 | 619 | 11.30 |
| 9 | 372.1 | 304 | 350 | 367 | 418 | 410 | 11.09 |
| 10 | 386.5 | 349 | 348 | 375 | 429 | 424 | 10.89 |
| Mean | **388.77** | 313 | 358 | 378 | 439 | 438 | **11.17** |
| Min | 323.7 | 248 | 292 | 315 | 375 | 372 | 10.80 |
| Max | 495.0 | 353 | 415 | 470 | 577 | 619 | 11.71 |
| SD | 46.25 | 34.3 | 37.6 | 44.4 | 56.9 | 71.2 | 0.30 |
| RMSE/Mean | **0.0350** | 0.03 | 0.03 | 0.03 | 0.04 | 0.04 | |

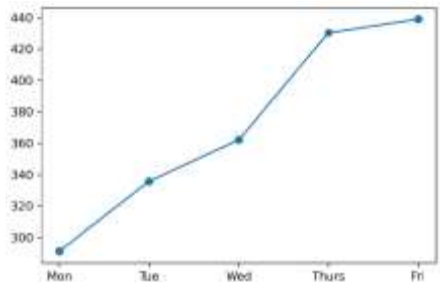

Fig. 10. Variation of RMSE with different days in a week for the LSTM#3 model (Round# 5 of Table V)

We observe that while the LSTM #3, the *univariate encoder-decoder Convolutional LSTM* is the fastest model, CNN #1- the univariate CNN model with the previous one week's data as the input, is found to be the most accurate model.

VI. CONCLUSION

In this paper, we have proposed five deep learning-based regression models for the prediction of NIFTY 50 index values. Our propositions included two CNN models and three LSTM models. The models were built, optimized, and then tested on the daily index values of NIFTY 50. While all the models exhibited high levels of accuracy in their forecasting performance, the univariate encoder-decoder convolutional LSTM with previous two weeks' data as its input, was found to be the most accurate model. However, in terms of execution speed, the univariate CNN model with previous one week's data as the input was found to be the fastest one. As a future scope of work, we plan to explore the possibility of designing generative adversarial networks (GANs)-based predictive models in order to further improve the forecasting accuracy.